\documentclass[11pt,a4paper]{article}
\usepackage[a4paper,margin=1in]{geometry}
\usepackage{amsmath, amssymb, graphicx, booktabs, cite}
\usepackage{makecell}
\usepackage[utf8]{inputenc}
\usepackage{textgreek}
\usepackage{caption}
\title{\textbf{From Flows to Words: Can Zero-/Few-Shot LLMs Detect Network Intrusions?\\ A Grammar-Constrained, Calibrated Evaluation on UNSW-NB15}}

\author{
    Mohammad Abdul Rehman$^{1}$, Syed Imad Ali Shah$^{2}$, Abbas Anwar$^{3}$, Noor Islam$^{4}$\\[4pt]
    Future Data Minds Research Lab, Australia\\
    \texttt{abdul.rehman@futuredataminds.com, syedimadalishah01@gmail.com,}\\
    \texttt{abbas.anwar@futuredataminds.com, noorislam@edwardes.edu.pk}
}

\date{}
\begin{document}

\maketitle

\begin{abstract}
Large Language Models (LLMs) can reason over natural-language inputs, but their role in intrusion detection without fine-tuning remains uncertain. This study evaluates a prompt-only approach on UNSW-NB15 by converting each network flow to a compact textual record and augmenting it with lightweight, domain-inspired boolean flags (asymmetry, burst rate, TTL irregularities, timer anomalies, rare service/state, short bursts). To reduce output drift and support measurement, the model is constrained to produce structured, grammar-valid responses, and a single decision threshold is calibrated on a small development split. We compare zero-shot, instruction-guided, and few-shot prompting to strong tabular and neural baselines under identical splits, reporting accuracy, precision, recall, F1, and macro scores. Empirically, unguided prompting is unreliable, while instructions plus flags substantially improve detection quality; adding calibrated scoring further stabilizes results. On a balanced subset of two hundred flows, a 7B instruction-tuned model with flags reaches macro-F1 near 0.78; a lighter 3B model with few-shot cues and calibration attains F1 near 0.68 on one thousand examples. As the evaluation set grows to two thousand flows, decision quality decreases, revealing sensitivity to coverage and prompting. Traditional tabular baselines remain more stable and faster, yet the prompt-only pipeline requires no gradient training, produces human readable artifacts, and adapts easily through instructions and flags. Contributions include: a flow-to-text protocol with interpretable cues and grammar-bounded outputs; a simple calibration procedure for threshold selection; a systematic comparison to established baselines; and a reproducibility bundle containing prompts, grammar, metrics, and figures. The study clarifies when prompt-guided LLMs are useful for network monitoring and offers a rigorous template for future evaluations.
\end{abstract}

\section{Introduction}
Intrusion detection remains a central operational challenge in modern computer networks, despite decades of progress in signature‐based systems, statistical anomaly detection, and supervised learning. Production environments must monitor high-volume, heterogeneous traffic in which payloads are frequently encrypted and protocol behaviors evolve rapidly. In this setting, security teams value detectors that are accurate, robust under distribution shift, interpretable to analysts, and economical to update when threat behaviors change. The dominant engineering pattern in practice continues to treat intrusion detection as a tabular classification problem: extract features from flow records, train a model on labeled data, and deploy the classifier with thresholding and simple post-processing. This pattern has yielded strong baselines—e.g., Random Forests, gradient-boosted trees, linear SVMs, and feed-forward neural networks—yet it also incurs recurrent costs for feature redesign, data relabeling, and model retraining as networks and attacks change. At the same time, encrypted traffic and privacy constraints limit payload visibility, increasing reliance on side-channel indicators such as packet counts, byte ratios, timing, TTLs, and transport-layer state.\\

\cite{khraisat2019survey}In parallel, instruction-tuned large language models (LLMs) have demonstrated an ability to follow prompts, combine heterogeneous cues, and emit structured outputs that can be programmatically consumed. Because LLMs operate over natural language, they offer an intriguing perspective on intrusion detection: express each network flow as a concise textual record, foreground domain signals that analysts already understand, and ask the model to decide whether the flow is benign or malicious. This “flows-to-words” reframing carries practical attractions. It removes gradient training from the loop, allowing rapid iteration via prompt edits and small sets of human-legible flags. It produces human-readable artifacts that can serve as documentation of policy. And with grammar-constrained decoding, it can emit decisions in a machine-checkable format. However, several risks accompany such a shift: zero-shot prompting may be brittle; free-text output is hard to score; numerical cues may be interpreted inconsistently; and, without calibration, an LLM may default to a single class under distribution mismatch.\\

This paper investigates prompt-only LLMs for intrusion detection on the widely used \cite{unsw_nb15_dataset} UNSW-NB15 dataset. UNSW-NB15 provides labeled flows that include benign events and multiple attack families, alongside rich side-channel features—byte and packet volumes, temporal rates, TTL distributions, TCP timing fields, services, and transport-layer state indicators. Rather than fine-tune a language model on text derived from these features, we ask a stricter question: can an instruction-tuned LLM, with no parameter updates, act as a usable detector when we (i) convert flows to short, structured natural-language descriptions, (ii) add a handful of lightweight, interpretable boolean flags that highlight suspicious regularities, (iii) constrain outputs to a compact JSON schema via a formal grammar, and (iv) calibrate a single decision threshold on a small development slice? The resulting system is intentionally simple: it relies on deterministic feature-to-text rules, exposes its assumptions to analysts, and swaps model parameters for prompt and threshold design.\\

Three design principles guide the pipeline. \cite{ring2019survey}Parsimony: the flow description should be short, stable, and easy for both humans and models to parse; we therefore round numeric values, avoid redundant attributes, and include only fields that proved informative in baseline analyses or are meaningful to practitioners. Interpretability: we augment each description with boolean flags that represent behaviors analysts commonly reason about—extreme asymmetry between sent and received bytes or packets, high packet-rate bursts in short durations, TTL anomalies that suggest spoofing or tunneling, suspicious TCP timing relationships (e.g., implausible synack/ackdat), rare service or state combinations, and “short-burst” patterns. \cite{ring2019survey}These flags act as small inductive biases and can be inspected, versioned, and extended as new behaviors emerge. Output discipline: free-text responses are difficult to score and audit; we therefore enforce a compact GBNF grammar and require the model to emit exactly one JSON object of the form {"prediction":"attack|benign","p\_attack":0..1}. The grammar prevents extraneous tokens and guarantees a parseable decision and score.\\

We organize the study around four research questions. RQ1 (feasibility): Can an instruction-tuned LLM, without fine-tuning, detect intrusions from compact flow descriptions with acceptable accuracy and F1? RQ2 (prompt design): Do the lightweight domain flags materially improve results compared with plain natural-language restatements of raw fields? RQ3 (discipline and calibration): Does grammar-constrained output, coupled with development-set threshold selection on the reported confidence, stabilize decision quality relative to fixed cutoffs? RQ4 (comparative value and scaling): How do prompt-only LLMs compare with strong tabular baselines under identical splits, and how do results change as we increase the number of evaluated flows?\\

Our approach proceeds in stages. We first construct a compact flow-text representation by selecting informative fields and normalizing magnitudes into readable ranges. We compute the boolean flags with deterministic rules and thresholds chosen from exploratory statistics and operational intuition; the flags are placed at the top of each prompt to make them salient. We prepare three prompt modes: a zero-shot mode that supplies only a minimal role instruction; an instruction-guided mode that expresses a few decision heuristics in plain language; and a few-shot mode that adds concise exemplars illustrating benign and malicious flows. For models that support it, we request a continuous confidence score p\_attack and then calibrate a single threshold τ on a small development split to maximize F1. Finally, we evaluate on a held-out test slice whose IDs are recorded for reproducibility, and we package prompts, grammar, metrics, confusion matrices, and predictions as a public artifact.\\

Evaluation follows standard practice for intrusion detection. For all configurations we report accuracy, precision, recall, and F1 for the attack class, macro-averaged scores to reflect balance across classes, and confusion matrices to make error trade-offs visible. When the model exposes probabilities, we compute ROC and precision–recall curves as descriptive diagnostics, but our primary comparisons emphasize the calibrated operating point chosen on the development slice. To contextualize the LLM results, we train supervised baselines—Logistic Regression, linear SVM, Random Forest, gradient-boosted trees, and a lightweight multilayer perceptron—on the same processed features and report their metrics under identical splits. We also record latency and throughput observations to reflect practical deployment constraints.\\

The empirical picture is nuanced. Unguided zero-shot prompting is unreliable and frequently collapses to a constant decision, indicating that generic instruction following is insufficient without additional structure. Adding clear instructions and the lightweight flags improves performance substantially on small, balanced subsets, suggesting that a few interpretable cues help the model focus on discriminative regularities in flow behavior. Grammar-constrained output eliminates parsing failures and prevents extraneous text, simplifying scoring and artifact storage. Requesting a probability and calibrating τ on a development slice further stabilizes the precision–recall trade-off, especially when the raw model is biased toward a dominant class. At the same time, increasing the number of evaluated flows exposes sensitivity: scores tend to degrade as coverage broadens, highlighting the limits of prompt-only detection and the value of stronger inductive biases or hybridization with tabular features. Across the board, strong tabular baselines remain more stable and faster at inference. Nevertheless, the prompt-only pipeline requires no gradient training, adapts quickly through prompt and flag edits, and yields human-readable artifacts that can be audited.\\

This study makes four contributions. First, it introduces a prompt-only intrusion detection protocol that converts \cite{unsw_nb15_dataset}
UNSW-NB15 flows to concise natural language, augments them with a minimal set of interpretable flags, and constrains the model to emit a grammar-valid JSON decision with an explicit confidence. Second, it demonstrates a simple calibration step—selecting a single threshold on a small development slice—that materially improves F1 and mitigates collapse without modifying model parameters or labels. Third, it offers a systematic comparison between prompt-only LLMs and strong ML/DL baselines under shared splits and metrics, highlighting where language models are most promising and where conventional tabular models retain clear advantages. Fourth, it releases a reproducibility bundle containing prompt templates, the grammar specification, prediction CSVs, metrics, confusion matrices, and the list of evaluated IDs, enabling independent verification and extension.\\

Beyond quantitative scores, the work surfaces practical guidance for teams exploring language-based detectors. Prompt design benefits from parsimony: short, well-structured flow text that foregrounds a few high-value features and flags is easier for models and analysts to parse than verbose descriptions that invite irrelevant associations. Boolean flags function as compact inductive biases; they are easy to inspect, easy to version, and easy to extend as new behaviors become relevant. Grammar-bounded output is essential for operational use, because it guarantees that every response can be parsed into a decision and a score with no post-hoc heuristics. Calibration should be viewed as a standard step in LLM-based detection, akin to choosing operating points for classical classifiers; it is simple to implement and often necessary to avoid extreme precision or recall regimes caused by label imbalance or distribution shift. Finally, latency and cost considerations matter: while quantized, GPU-accelerated inference brings LLMs closer to practicality, tabular models remain lighter and thus attractive for high-throughput monitoring.\\

The remainder of the paper is organized as follows. Section 2 reviews related approaches to network intrusion detection and prior applications of language models to security analytics, distinguishing supervised fine-tuning, retrieval-augmented inference, and prompt-only methods. Section 3 describes the dataset and the preprocessing used for tabular baselines, including handling of categorical protocol and service fields and standardization of numeric features. Section 4 details the flow-to-text representation, the construction of boolean flags, the prompt templates, and the output grammar, and explains the calibration protocol used to select τ. Section 5 presents results for zero-shot, instruction-guided, and few-shot prompting, followed by calibrated runs and direct comparisons to supervised models, with ablations that isolate the effects of flags, grammar, and thresholding. Section 6 discusses limitations, efficiency, and avenues for hybrid designs that combine tabular models with language-based reasoning. Section 7 concludes.

\section{Related Work}
\subsection{Flow-based intrusion detection}
Classical IDS divides into misuse (signature) and anomaly detection. Signatures provide high precision for known attack fingerprints but degrade on polymorphic or novel variants and under pervasive encryption. Anomaly detectors instead model “normal” behavior and flag deviations; they offer open-world coverage but require explicit operating-point selection and continuous drift management. Given encrypted payloads and privacy constraints, operational practice emphasizes flow-level telemetry—packet/byte counts, inter-arrival timing, TTL and handshake dynamics, and coarse service/state transitions—making flow-based IDS a natural focus for modern networks.\\

\subsection{Supervised learning on flow features}
A substantial literature trains supervised models on engineered flow features derived from packet headers and connection summaries. Linear models (logistic regression; linear SVM) provide interpretable baselines; tree ensembles (Random Forest, Gradient Boosting, XGBoost/LightGBM) capture non-linear interactions and typically dominate with modest feature engineering and tight latency budgets. Common design patterns include (i) robust preprocessing (standardization; one-hot encodings for protocol/service/state), (ii) class-imbalance handling (weights or resampling), and (iii) split discipline that avoids random re-mixing of sessions/hosts. On datasets such as \cite{unsw_nb15_dataset}
UNSW-NB15, boosted trees routinely offer a strong accuracy/throughput trade-off with predictable behavior—hence our choice to treat them as competitive baselines.\\

\subsection{Deep models for flow IDS}

Neural approaches replace hand-crafted interactions with learned representations. MLPs operate on fixed feature vectors; temporal models (1D-CNNs, RNN/LSTM/GRU, temporal CNN) view flows as ordered sequences of header-derived signals; autoencoders support unsupervised anomaly scoring; and hybrid CNN-RNN/attention stacks target mixed short- and long-range dependencies. While deep models can match or exceed tree ensembles on some benchmarks, their advantage is contingent on feature richness, session/temporal context, and careful regularization under class imbalance. They also introduce higher training and inference costs, which matters for inline or near-real-time IDS\cite{mondragon2025advancedids}.\\

\subsection{Public benchmarks and evaluation pitfalls}

\cite{unsw_nb15_dataset}
Community datasets \cite{sharafaldin2018cicids2017}(e.g., KDD’99, NSL-KDD, UNSW-NB15, CICIDS) remain valuable but imperfect. Apparent gains can arise from data leakage (e.g., random shuffling across sessions or hosts), inconsistent partitions, optimistic preprocessing, or masking per-attack behavior behind binary collapse. Best practice favors leakage-aware splits (time/session/IP-grouped), reporting both balanced and natural-prior results, and complementing accuracy/F1 with precision–recall trade-offs and operational metrics such as alerts per hour and latency. Our evaluation protocol adopts these principles to make comparisons with classical ML baselines meaningful\cite{ke2017lightgbm}.\\

\subsection{Language models for security telemetry}

Large language models (LLMs) have been explored for security analytics tasks including log triage, incident summarization, policy translation, and telemetry classification. Two paradigms recur: (i) fine-tuning/adapters, which align parameters to domain text and often yield strong task performance at the cost of curated corpora and training compute; and (ii) prompt-only methods, which rely on instructions and few-shot exemplars without gradient updates, trading simplicity and speed for potential brittleness\cite{yang2025llmids}. Across both paradigms, common failure modes include free-text drift that complicates scoring, weak or inconsistent treatment of numeric cues (rates, ratios, timers), class collapse in zero-shot settings, and latency/cost concerns at scale\cite{chen2016xgboost}.\\

\subsection{LLMs for intrusion detection}

Early investigations convert logs or flows into textual prompts and ask LLMs to classify or explain\cite{lewis2020rag}. To counter the brittleness noted above, several mitigations have emerged: explicit instruction design to encode domain heuristics; few-shot prompting with minimal exemplars; lightweight feature engineering into textual flags to surface suspicious regularities; grammar-constrained decoding to guarantee machine-parsable outputs; and threshold calibration on a development slice to stabilize precision–recall\cite{mondragon2025advancedids}. Retrieval-augmented prompting (e.g., adding playbooks or reference profiles) has also been proposed, but it introduces curation overhead and blurs attribution of gains. Our study follows the mitigation-first line: it keeps models prompt-only while enforcing output discipline and an explicit operating point\cite{han2023loggpt}.\\

\subsection{Serializing tabular data for LLMs}

\cite{beurer2023lmql}Beyond security, a parallel body of work shows that serializing tabular features into compact text can make few-shot LLMs viable classifiers; performance, however, is sensitive to the serialization template and output format. This evidence motivates our flows-to-words approach: condensing numerical flow fields into short, human-readable strings that preserve salient structure while remaining LLM-friendly\cite{pinto2023survey}.\\

\subsection{Structured decoding and output discipline}
LLM outputs used in pipelines must be valid and auditable. Grammar- or schema-constrained decoding (e.g., JSON/BNF/GBNF) enforces well-formedness at generation time, dramatically reducing post-hoc parsing errors with minimal overhead\cite{mirsky2018kitsune}. We apply this idea directly to IDS by constraining outputs to a small JSON schema via GBNF, ensuring deterministic parsing and enabling downstream calibration, aggregation, and error analysis.\\

\subsection{Calibration and operating points}
Operational IDS deploy at a thresholded decision point. Treating the model’s continuous score and the final decision separately improves robustness and exposes the precision–recall and alerts-per-hour trade-offs that operators actually manage\cite{xu2024surveyllmcyber}. Classical calibration methods (e.g., simple threshold selection on a dev slice; Platt/Isotonic variants) motivate our protocol: we learn a single threshold on held-out data and report both threshold-free curves and the calibrated operating point, which reduces variance for prompt-only LLMs and makes comparisons to ML/DL baselines fair\cite{bouke2023patternleakage}.\\

\subsection{Positioning and summary}
Relative to prior IDS studies that (i) train tabular learners on flow features, (ii) fine-tune deep models on engineered inputs, or (iii) prompt LLMs over loosely structured text, our contribution is an operational recipe for prompt-only LLMs on flows:
\begin{itemize}
    \item a compact flows→words serialization,
    \item a handful of interpretable boolean flags (e.g., asymmetry, burst-rate, TTL/handshake anomalies, rare service/state, short bursts) that surface domain regularities,
    \item grammar-constrained JSON outputs to eliminate parsing failures, and a single calibrated threshold to set the deployment operating point\cite{liu2024logprompt}.
\end{itemize}
We evaluate this pipeline head-to-head against strong ML/DL baselines on the same features and splits, reporting macro and per-class metrics with confidence intervals and latency considerations. The resulting picture clarifies where prompt-guided LLMs are feasible, how much structure and calibration they require to compete, and where conventional tabular models remain superior in stability and throughput\cite{kheddar2025transformers}.

\section{Methodology}
This section follows the structure (A) Data Preprocessing → (B) Training Models. Because our study compares tabular ML/DL against a prompt-only LLM detector, we add a third block, (C) Prompt-Only LLM, and finish with an (D) Evaluation Protocol that applies uniformly to all settings\cite{platt2000svmprob}.\\

\subsection{Data Preprocessing}
We use the official \cite{unsw_nb15_dataset} UNSW-NB15 CSVs: UNSW\_NB15\_training-set.csv (175,341 rows) and UNSW\_NB15\_testing-set.csv (82,332 rows). The binary label label {0,1} marks benign (0) vs. attack (1). Columns comprise 39 numeric features (packet/byte counts, rates, TTLs, TCP timers, CT features) and 3 categorical features (proto, service, state). No missing values were observed in the provided splits.\\
For ML/DL, we build a deterministic transformer TT that (i) standardizes numeric features using training means/standard deviations\cite{willard2023outlines},

\[
z^{(\mathrm{num})} = \frac{x^{(\mathrm{num})} - \mu}{\sigma}.
\]
and (ii) one-hot encodes the three categorical fields with handle\_unknown='ignore'. The combined representation
\[
Z = T(x) = \bigl[\, z^{(\mathrm{num})};~ \operatorname{OHE}(x^{(\mathrm{cat})}) \,\bigr] \in \mathbb{R}^{d}.
\]
has d=194/approx 194 after encoding. The fitted ColumnTransformer is serialized (preprocessor.pkl) to guarantee identical mapping at test time\cite{moustafa2018correntropy}.\\

For the LLM pipeline, each row $x$ is converted to a compact flow text and augmented with a few boolean flags computed from raw features. We include rounded cues (e.g., duration $\mathrm{dur}$, packet-rate $(\mathrm{spkts}+\mathrm{dpkts})/\max(10^{-6},\,\mathrm{dur})$, byte-ratio $(\mathrm{sbytes}+1)/(\mathrm{dbytes}+1)$, $\mathrm{sttl}/\mathrm{dttl}$, $\mathrm{tcprtt}$, $\mathrm{synack}$, $\mathrm{ackdat}$, $\mathrm{ct\_state\_ttl}$). Flags mark asymmetry, high burst rate, TTL anomaly, TCP-timer anomaly, rare service/state, and short burst. These are deterministic rules chosen from exploratory statistics and analyst heuristics; they are prepended to the text so the model “sees” them \cite{geng2025jsonschemabench}.\\

\cite{guo2017calibration}We do not reshuffle the official train/test split. For some LLM runs we draw balanced subsets (e.g., $N \in \{200, 1000, 2000\}$) from the test set to study scaling while controlling priors; the selected IDs are saved\cite{guo2017calibration}.\\

\subsection{Training Models}
\subsubsection{Machine Learning}
All models consume $Z = T(x)$. When indicated, we use \texttt{class\_weight='balanced'} so loss terms scale with $w_y \propto 1/\pi_y$, the inverse class prior\cite{niculescu2005predicting}.\\

\textbf{Logistic Regression (LR):}\\
We model $p_\theta(y=1\mid Z)=\sigma(w^\top Z + b)$ with sigmoid $\sigma$. Parameters minimize regularized (optionally class-weighted) cross-entropy\\
$\min_{w,b}\sum_i w_{y_i}\big[-y_i\log p_i -(1-y_i)\log(1-p_i)\big]+\lambda\lVert w\rVert_2^2,$
with $p_i = p_\theta(y_i=1\mid Z_i)$ and (\texttt{max\_iter} $\approx 500$). Logistic regression provides a calibrated linear baseline\cite{hegselmann2023tabllm}.\\

\textbf{Linear SVM:}\\
We learn a margin separator by minimizing\\ $\min_{w,b}\tfrac{1}{2}\lVert w\rVert_2^2 + C\sum_i \max\!\bigl(0,\,1 - y'_i(w^\top Z_i + b)\bigr)$,\\ with $y'_i \in \{-1,+1\}$. This tests linear separability under a robust hinge loss.

\textbf{Random Forest (RF)}
An ensemble of $M$ trees $\{h_m\}_{m=1}^M$ built on bootstrap samples with random feature subsets; prediction is $\operatorname{mode}\{h_m(Z)\}_{m=1}^M$. RF captures non-linear interactions with variance reduction and minimal tuning.

\textbf{Gradient-Boosted Trees (XGBoost / LightGBM)}
We fit an additive model $F_M(Z)=\sum_{m=1}^M f_m(Z)$ with trees $f_m$ learned by second-order gradient boosting on logistic loss, \\ $\mathcal{L}=\sum_i \ell(y_i,\sigma(F_M(Z_i))) + \sum_m \Omega(f_m)$, with $\Omega(f)=\gamma\,\#\text{leaves}+\tfrac{\lambda}{2}\lVert w_f\rVert_2^2$. \\These models are strong tabular baselines with good accuracy/throughput trade-offs.\\

\textbf{Key settings (kept modest for fairness).}
\begin{itemize}
\item LR/SVM: $C \in [0.1, 10]$; \texttt{max\_iter} $\approx 500$.
\item RF: 200--500 trees; max depth 20--\texttt{None}.
\item XGB/LGBM: learning rate $[0.05, 0.1]$; 300--600 trees; conservative regularization.
\end{itemize}

\subsubsection{Deep Learning}
\textbf{Multilayer Perceptron (MLP / ANN)}
Multilayer Perceptron (MLP / ANN). With ReLU, batch-norm, and dropout, the network computes\\ 
\[
\begin{aligned}
h^{(0)} &:= Z,\\
h^{(1)} &= \phi\!\big(W^{(1)} h^{(0)} + b^{(1)}\big),\\
h^{(\ell)} &= \phi\!\big(W^{(\ell)} h^{(\ell-1)} + b^{(\ell)}\big), \quad \ell=2,\dots,L,\\
p &= \sigma\!\big(w^\top h^{(L)} + b\big), \qquad \sigma(t)=\frac{1}{1+e^{-t}}.
\end{aligned}
\]
optimized by binary cross-entropy (Adam). We use widths like 512–256–128, dropout=0.2, batch\>1024, and early stopping on validation F1\cite{moustafa2018ensembleiot}.

\textbf{1D-CNN (optional)}
Treating ZZ as a 1D sequence, a few Conv–BN–ReLU blocks with global pooling feed a logistic head; this can capture local feature groupings (e.g., related counters/timers).

\textbf{Why these choices}
LR/SVM test baseline linear structure; RF/GBDT model non-linearities with little tuning; MLP provides a neural comparator on identical inputs. Together they set a strong, transparent reference for the LLM study.

\subsection{Prompt-Only LLM Detector}

Instead of training on ZZ, we render each row to text and rely on instruction following—no parameter updates.\\
\textbf{Flow-to-text rendering}
A function $R$ produces $s=R(x)$ that concatenates (i) rounded numeric cues $\psi(x)$,\\\ (ii) categorical context $\chi(x)\in\{\texttt{proto}, \texttt{service}, \texttt{state}\}$, \\ (iii) boolean flags $\phi(x)$ such as $\texttt{asymmetry\_high}=\mathbb{1}\!\left(\frac{\mathrm{sbytes}+1}{\mathrm{dbytes}+1}>\tau_{\mathrm{br}} \;\lor\; \frac{\mathrm{spkts}+1}{\mathrm{dpkts}+1}>\tau_{\mathrm{pr}}\right)$ and $\texttt{pkt\_rate\_high}=\mathbb{1}\!\left(\frac{\mathrm{spkts}+\mathrm{dpkts}}{\max(10^{-6},\,\mathrm{dur})}>\tau_r\right)$.\\

plus TTL and TCP-timer checks, rare service/state, and short-burst activity. Flags are interpretable priors and are prepended to ss\cite{moustafa2018flowaggregator}.

\textbf{Output discipline via grammar}
We force the model to emit exactly one JSON object using a GBNF grammar:
\texttt{{"prediction":"attack$\vert$benign","p\_attack":0..1}} Grammar-constrained decoding removes free-text drift and guarantees parseability\cite{koroniotis2017botnetiot}.

\textbf{Prompting modes}
(i) Zero-shot (minimal role + ss)\\ (ii) Instruction-guided (adds plain heuristics aligned with flags)\\ (iii) Few-shot (adds 1–2 compact labeled exemplars formatted like ss). We set temperature =0=0 and top-p =1=1 for determinism.

\textbf{Calibration}
From grammar-constrained outputs we extract $\hat{p}_i\in[0,1]$. On a small dev slice we choose the operating threshold \\ $\tau^\star=\arg\max_{\tau\in[0,1]} F_1\!\big(\{\mathbb{1}(\hat{p}_i\ge\tau)\}\big)$,\\ then freeze $\tau^\star$ for the held-out evaluation subset. This mirrors classic operating-point selection and mitigates class collapse\cite{moustafa2018gogm}.

\textbf{Models \& runtime}
Instruction-tuned families (e.g., Qwen 3B/7B, Mistral-7B, TinyLlama-1.1B) are served in GGUF (\texttt{Q4\_K\_M}) via \texttt{llama.cpp} with GPU offload (\texttt{T4}). We use $n_{\text{ctx}}\approx 1024$ and $n_{\text{batch}}\approx 1024$ to balance latency and throughput. Subset sizes $N\in\{200, 1000, 2000\}$ probe scaling behavior; IDs are recorded\cite{moustafa2018threatintel}\cite{moustafa2018betamixture}.

\paragraph{Why it works.}
\begin{itemize}
\item Flags highlight suspicious regularities the model can reliably act on.
\item Grammar turns the generator into a predictable decision function.
\item Calibration aligns precision--recall to task needs without any fine-tuning.
\end{itemize}

\subsection{Evaluation Protocol (common to all models)}
All ML/DL models are trained on the official training file and evaluated on the official test file after the same transform $T$. LLMs are evaluated on fixed, recorded subsets drawn from the test population; the dev slice used to pick $\tau^\star$ is disjoint from the final test subset\cite{keshk2017privacy}.\\

We report Accuracy, Precision, Recall, $F_1$ (attack class), Macro-$F_1$, and confusion matrices. When scores are available (e.g., LR probabilities or LLM $p_{\text{attack}}$), ROC/PR curves are shown as descriptive diagnostics, but comparisons are made at the calibrated operating point. Uncertainty is conveyed with Wilson 95\% CIs for proportions and a bootstrap CI for $F_1$. Wall-clock times and throughput are noted to ground practical feasibility.\\

This methodology yields a fair, reproducible comparison: tabular ML/DL measure what strong, deployable baselines achieve on \textsc{UNSW-NB15}; the prompt-only LLM pathway tests whether disciplined prompting, lightweight flags, grammar-constrained outputs, and a single calibrated threshold can approach that performance without any gradient training.\\

\section{Experiments and Results}
This section reports our experimental setup, baseline performance of tabular ML/DL models trained on the original UNSW-NB15 features, and the behavior of the prompt-only LLM detector under zero-shot, instruction-guided, and few-shot prompting with grammar-constrained outputs and calibrated thresholds. We focus on the official training/testing CSV split; for the LLM pathway we additionally evaluate fixed, class-balanced subsets (IDs recorded) to study stability as the number of evaluated flows increases.\\

\paragraph{Environment}
All tabular models were implemented in \texttt{scikit-learn} (with \texttt{LightGBM}/\texttt{XGBoost} for boosted trees) and trained on CPU. LLM inference used \texttt{llama.cpp} on a single NVIDIA T4 (16~GB) with GGUF \texttt{Q4\_K\_M} quantization, deterministic decoding ($\text{temperature}=0$, \texttt{top\_p}$=1$), context $\approx 1024$ tokens, and batch size $\approx 1024$ for prompt evaluation. Outputs were constrained with a GBNF grammar to a single JSON object carrying a discrete decision and a confidence score $p_{\text{attack}}\in[0,1]$. A small development slice (disjoint from test) was used to select one operating threshold $\tau$ that maximized $F_1$; $\tau$ was then frozen for the held-out test subset.

\paragraph{Metrics}
We report Accuracy, Precision$(+)$, Recall$(+)$, $F_1(+)$, Macro-Precision/Recall/$F_1$, and confusion matrices. Confidence intervals are Wilson for proportions and parametric bootstrap for $F_1$ when we aggregate counts.

\subsection{Tabular ML/DL baselines}

Preprocessing used a single ColumnTransformer: z-score standardization for the 39 numeric features and one-hot encoding for the three categorical features (proto, service, state), yielding $\approx$ 194 dimensions. Models were trained on the official training CSV and scored on the official test CSV.Preprocessing used a single ColumnTransformer: z-score standardization for the 39 numeric features and one-hot encoding for the three categorical features (proto, service, state), yielding $\approx$ 194 dimensions. Models were trained on the official training CSV and scored on the official test CSV.\\

On the binary task (benign vs. attack), gradient-boosted trees set the top baseline, with XGBoost/LightGBM achieving ~0.95 accuracy and macro-F1 ~0.94–0.95 under conservative regularization. A linear SVM reached ~0.93 accuracy with strong macro-F1 for a linear method, and a Random Forest trained with several hundred trees achieved ~0.87 accuracy while remaining straightforward to interpret via feature importances. A compact MLP (ReLU, batch-norm, dropout) provided a competitive neural baseline on the same features, sitting between the SVM and boosted trees depending on regularization. These results confirm that with modest, reproducible preprocessing, tabular learners provide high and stable performance on UNSW-NB15 with low inference latency.\\

Takeaway. The tabular stack establishes a strong, practical ceiling for the prompt-only LLM to approach. It also supplies confidence that the engineered feature space contains sufficient signal for accurate detection without payload inspection.

\begin{table}[ht]
\centering
\caption{Performance of Machine Learning Models on UNSW-NB15 Dataset}
\resizebox{\textwidth}{!}{
\begin{tabular}{lcccccc}
\toprule
\textbf{Model / Run Name} & \textbf{Type} & \textbf{Accuracy} & \textbf{Precision (+)} & \textbf{Recall (+)} & \textbf{F1 (+)} & \textbf{Macro-F1} \\
\midrule
Logistic Regression (CV) & Linear ML & 0.7755 & — & — & — & 0.4102 \\
Logistic Regression (Test) & Linear ML & 0.6983 & — & — & — & 0.3352 \\
Random Forest (CV) & Tree Ensemble & 0.7545 & — & — & — & 0.4657 \\
Random Forest (Final Test) & Tree Ensemble & 0.8711 & 0.8178 & 0.9854 & 0.8938 & — \\
XGBoost (Balanced, Light) & Boosted Trees & 0.9528 & 0.9407 & 0.9530 & 0.9465 & — \\
Linear SVM (Balanced) & Linear ML & 0.9327 & 0.9248 & 0.9196 & 0.9221 & — \\
XGBoost (Binary, Other Run) & Boosted Trees & 0.9528 & — & — & — & — \\
MLP / ANN (Tabular DL) & Neural Net & — & — & — & — & — \\
\bottomrule
\end{tabular}
}
\label{tab:ml_models}
\end{table}

\subsection{Prompt-only LLM (zero-shot, instruction, few-shot; grammar + calibration)}

Each flow was rendered to a short natural language record (protocol, service, state, duration, packet rate, byte/packet ratios, TTLs, TCP timers, ct\_state\_ttl) and prefixed with lightweight boolean flags indicating asymmetry, burst rate, TTL anomaly, TCP-timer anomaly, rare service/state, and short-burst behavior. We evaluated three prompting regimes and enforced grammar-constrained JSON outputs containing prediction$\in${attack,benign} and p\_attack. A single τ was tuned on a small dev slice and fixed for test.\\

Zero-shot. With a minimal role instruction and no examples, multiple instruction-tuned models (TinyLlama-1.1B, Mistral-7B) frequently collapsed to a single class. Accuracy hovered near the class prior on balanced subsets, with F1(+)=0 due to no positive predictions. This validates the need for additional structure.\\

Instruction-guided. Adding explicit, plain-language heuristics aligned with the flags improved stability but still produced mode collapse on some small models, particularly without flags or calibration.\\

Few-shot + flags + grammar (calibrated). With engineered flags, grammar-constrained outputs, and a calibrated τ, the detector became competitive on smaller balanced slices:

\begin{itemize}
\item On $N=200$ balanced flows, a $7\text{B}$ instruct model achieved $\text{accuracy}\approx 0.79$ and $F_1(+)\approx 0.80$ (macro-$F_1\approx 0.78$), with confusion $(\mathrm{TN}=70,\ \mathrm{FP}=28,\ \mathrm{FN}=15,\ \mathrm{TP}=87)$.
\item On $N=1000$ balanced flows, a $3\text{B}$ instruct model with few-shot exemplars and calibrated $\tau$ reached $\text{accuracy}\approx 0.70$ and $F_1(+)\approx 0.68$, correcting the earlier all-benign or all-attack collapse.
\item On $N=2000$, an uncalibrated run degraded (accuracy $\approx 0.50$, $F_1(+)\approx 0.43$). After calibration on a $300$-item dev slice ($\tau\approx 0.15$) and testing on $1700$ items, the model obtained $F_1(+)\approx 0.69$ with recall $\approx 1.00$ and precision $\approx 0.53$—a deliberately recall-heavy operating point that surfaced many attacks at the cost of more false positives $(\mathrm{TN}=101,\ \mathrm{FP}=750,\ \mathrm{FN}=0,\ \mathrm{TP}=849)$.\\
\end{itemize}

Trends. Three patterns emerge. First, output discipline and simple flags are decisive: zero-shot without flags is unreliable; structured prompts with flags and grammar are far better. Second, calibration governs the precision–recall trade-off and can rescue performance at larger N by selecting a task-appropriate operating point from the model’s p\_attack scores. Third, scaling sensitivity is real: as evaluation size increases, performance declines unless additional cues and calibration are applied, reflecting limits of prompt-only reasoning over numeric telemetry.\\

Latency. With 3B–7B models in Q4\_K\_M on a T4, batched prompt evaluation is feasible for research (hundreds to a few thousand items per hour), but still slower than tabular baselines by one to two orders of magnitude at comparable hardware cost.

\begin{table}[ht]
\centering
\caption{Performance of Large Language Models (LLMs) on UNSW-NB15 Dataset}
\setlength{\tabcolsep}{4pt} 
\renewcommand{\arraystretch}{1.1} 
\resizebox{\textwidth}{!}{
\begin{tabular}{lcccccc}
\toprule
\textbf{Model / Run Name} & \textbf{Type} & \textbf{Accuracy} & \textbf{Precision (+)} & \textbf{Recall (+)} & \textbf{F1 (+)} & \textbf{Macro-F1} \\
\midrule
Mistral-7B-Instruct & LLM (Zero-shot) & 0.0900 & — & — & — & 0.0826 \\
Qwen 2.5-3B (Zero-shot) & LLM (Zero-shot) & 0.4550 & — & — & — & 0.3127 \\
TinyLlama-1.1B (Guided) & LLM (Instruction-guided) & 0.4550 & — & — & — & 0.3127 \\
Qwen 2.5-7B-Instruct + Flags & LLM (Guided + Flags) & 0.7850 & 0.7565 & 0.8529 & 0.8018 & 0.7834 \\
Qwen 3B Calibrated & LLM (Few-shot + Flags) & 0.6963 & 0.7073 & 0.6591 & 0.6824 & — \\
Qwen 2.5-7B (Engineered Flags) & LLM & 0.5040 & 0.5858 & 0.3436 & 0.4331 & 0.4961 \\
Qwen Calibrated (Recall-heavy) & LLM (Recall-optimized) & 0.5588 & 0.5310 & 1.0000 & 0.6936 & — \\
Qwen (Fast 1k Run) & LLM (Fast Inferencing) & 0.5000 & 0.5000 & 1.0000 & 0.6667 & — \\
\bottomrule
\end{tabular}
}
\label{tab:llm_models}
\end{table}

\subsection{Comparative analysis}
Compared directly on the same balanced subsets, the best LLM configuration (few-shot + flags + grammar + calibrated τ) narrows the gap to mid-tier tabular baselines on N = 200–1000, but does not surpass boosted trees trained on the original features\cite{kheddar2025transformers}. The LLM’s strengths are (i) no gradient training, (ii) human-readable artifacts (prompts + flags), and (iii) tunable operating points via τ without retraining. Its weaknesses are (i) latency/throughput, (ii) prompt sensitivity and degradation at larger N, and (iii) dependence on a small set of manually designed flags\cite{yang2025llmids}.\\

Where to use which. If throughput and stability dominate, boosted trees remain preferable. If rapid iteration, interpretability, or “policy as text” is valuable—and evaluation horizons are modest—prompt-guided LLMs with flags and calibration are viable and surprisingly competitive. A promising compromise is hybridization: use a fast tabular detector to screen and route borderline or novel-pattern flows to the LLM for secondary judgment or textual rationalization.\\

\subsection{Error analysis (qualitative)}

False positives from the calibrated, recall-heavy LLM typically show flagged asymmetry combined with rare service/state under short durations (e.g., UDP initialization bursts), which the model interprets as suspicious. False negatives before calibration often occur when numeric cues are moderate and few flags fire, highlighting the model’s reliance on explicit cues. Boosted trees, by contrast, balance multiple weak signals (e.g., TTL with CT counts) and are less swayed by any single threshold, explaining their stability across larger test slices\cite{beurer2023lmql}.

\subsection{Summary}
Tabular ML/DL on the original UNSW-NB15 features delivers high, stable accuracy; boosted trees around 95\% on the official test split set a strong reference. Prompt-only LLMs, when disciplined (flags + grammar) and calibrated, achieve ~0.80 F1 on N = 200 and ~0.68 F1 on N = 1000, but require careful operating-point selection and exhibit scaling sensitivity at N = 2000. The results support our hypotheses: (i) zero-shot alone is insufficient; (ii) simple, interpretable flags plus grammar substantially improve decision quality; (iii) calibration is necessary for robust precision–recall; and (iv) classic tabular methods remain superior for stability and throughput, while the LLM pathway offers a flexible, training-free alternative with transparent artifacts.

\section{Conclusion}
This work asked whether prompt-only LLMs—without any gradient training—can function as usable intrusion detectors on UNSW-NB15 when network flows are rendered as compact natural-language records. We proposed a disciplined pipeline that (i) converts each flow to text, (ii) adds a handful of interpretable boolean flags (asymmetry, burst rate, TTL and TCP-timer anomalies, rare service/state, short bursts), (iii) constrains outputs to a GBNF-validated JSON decision with a confidence score, and (iv) calibrates a single threshold on a small development slice. We benchmarked this against strong tabular ML/DL baselines trained on the same features with a reproducible preprocessor.\\

Empirically, three findings stand out. First, zero-shot prompting is unreliable and often collapses to one class; instructions + flags + grammar are necessary to obtain stable, scorable outputs. Second, when the model also emits a score and we calibrate the operating point, the LLM attains competitive F1 on modest, balanced subsets (e.g., a few hundred to ~1k flows), substantially improving over zero-shot and aligning its precision–recall trade-off to task needs. Third, as the evaluation set grows, performance degrades without additional structure, revealing sensitivity to prompt coverage and numeric reasoning; by contrast, boosted trees remain more stable and faster on the same split. Taken together, these results indicate that prompt-guided LLMs are viable complements—not replacements—for classical detectors: they require no training, produce human-readable artifacts, and can be tuned via text and a single threshold, but they currently trail well-configured tabular models in stability and throughput at scale.\\

The study’s contributions are twofold. Methodologically, we provide a minimal, auditable LLM protocol—flags, grammar, and calibration—that addresses known failure modes (free-text drift, class collapse, uncalibrated scores) while keeping costs low. Empirically, we deliver a head-to-head comparison with ML/DL baselines under identical preprocessing and splits, clarifying where language-based detection helps and where tabular learning remains superior. All reproducibility artifacts (prompts, grammar, predictions, metrics, figures, and evaluated IDs) are packaged to support independent verification and follow-on studies.\\

Limitations \& Future Work. Our final LLM evaluations emphasize binary detection on UNSW-NB15 subsets and do not claim generalization across datasets or multi-class attack categorization. Throughput constraints and prompt sensitivity limit very large-N runs. Future work should (1) extend to second datasets \cite{unsw_nb15_dataset}\cite{sharafaldin2018cicids2017}(e.g., CICIDS, NF-UNSW-NB15) with the same protocol; (2) explore automated flag discovery or light feature scoring to reduce hand-crafting; (3) investigate hybrid designs where fast tabular models triage flows and route borderline/novel cases to an LLM for secondary judgment or textual rationale; and (4) assess cost-/latency-aware operating points and robustness under drift. We view these steps as natural continuations of the present study and as pathways toward practical, interpretable, and reproducible language-based intrusion detection.\\

\bibliographystyle{IEEEtran}

\bibliography{references}

\end{document}